\documentclass[conference]{IEEEtran}
\IEEEoverridecommandlockouts

\usepackage{cite}
\usepackage{hyperref}
\usepackage{amsmath,amssymb,amsfonts}
\usepackage{algorithm}
\usepackage{algpseudocode}
\usepackage{graphicx}
\usepackage{listings} 
\usepackage{xcolor}
\usepackage{grffile} 
\usepackage{textcomp}
\usepackage[table, svgnames, dvipsnames]{xcolor}
\usepackage{verbatim}
\usepackage{braket}
\usepackage{subcaption}
\usepackage{placeins}
\usepackage{multirow}
\usepackage{booktabs}
\usepackage{makecell}
\usepackage{amsmath,amssymb}
\usepackage{tikz}
\usetikzlibrary{arrows.meta,positioning,fit,calc,quotes}
\definecolor{darkgreen}{RGB}{0,100,0}

\def\BibTeX{{\rm B\kern-.05em{\sc i\kern-.025em b}\kern-.08em
    T\kern-.1667em\lower.7ex\hbox{E}\kern-.125emX}}

\begin{document}

\title{ Efficient Transpilation of OpenQASM 3.0 Dynamic Circuits to CUDA-Q: Performance and Expressiveness Advantages
}
\author{
\IEEEauthorblockN{
Vinooth Kulkarni,
Jaehyun Lee,
Adam Hutchings,
Anas Albahri,
Jai Nana,
Shuai Xu,
Vipin Chaudhary
}
\IEEEauthorblockA{
Dept. of Computer and Data Sciences, Case Western Reserve University\\
Cleveland, OH, USA\\
\{vxk285,jxl1646,ash160,ama456,jxn398,sxx214,vxc204\}@case.edu
}
\thanks{This research was supported in part by NSF Awards 2216923 and 2117439.}
}
\maketitle

\begin{abstract}
 
\end{abstract}

Dynamic quantum circuits with mid-circuit measurement and classical feedforward are essential for near-term algorithms such as error mitigation, adaptive phase estimation, and Variational Quantum Eigensolvers (VQE), yet transpiling these programs across frameworks remains challenging due to inconsistent support for control flow and measurement semantics. We present a transpilation pipeline that converts OpenQASM 3.0 programs with classical control structures (conditionals and bounded loops) into optimized CUDA‑Q C++ kernels, leveraging CUDA‑Q’s native mid-circuit measurement and host-language control flow to translate dynamic patterns without static circuit expansion. Our open-source framework is validated on comprehensive test suites derived from IBM Quantum’s classical feedforward guide—including conditional reset, if–else branching, multi-bit predicates, and sequential feedforward—and on VQE-style parameterized circuits with runtime parameter optimization. Experiments show that the resulting CUDA‑Q kernels reduce circuit depth by avoiding branch duplication, improve execution efficiency via low-latency classical feedback, and enhance code readability by directly mapping OpenQASM 3.0 control structures to C++ control flow, thereby bridging OpenQASM 3.0’s portable circuit specification with CUDA‑Q’s performance-oriented execution model for NISQ-era applications requiring dynamic circuit capabilities.


\begin{figure*}[t!]
    \centering

    \begin{subfigure}[t]{0.98\textwidth}
        \centering
        \includegraphics[width=\linewidth]{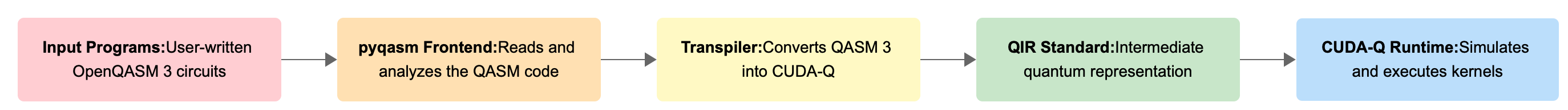}
        \caption{High-level ecosystem integrating user-written OpenQASM 3 circuits, the pyqasm frontend, and the CUDA-Q runtime.}
        \label{fig:ecosystem}
    \end{subfigure}

    \vspace{0.6em}

    \begin{subfigure}[t]{0.485\textwidth}
        \centering
        \includegraphics[width=\linewidth]{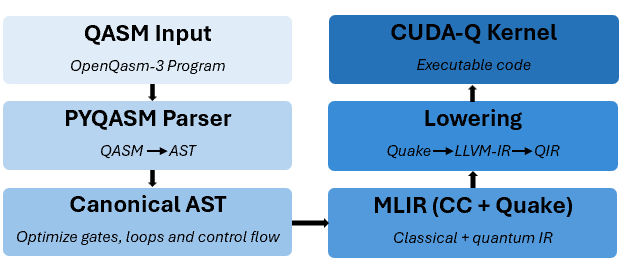}
        \caption{Main QASM $\to$ CUDA-Q compilation pipeline.}
        \label{fig:compilation_pipe}
    \end{subfigure}
    \hfill
    \begin{subfigure}[t]{0.485\textwidth}
        \centering
        \includegraphics[width=\linewidth]{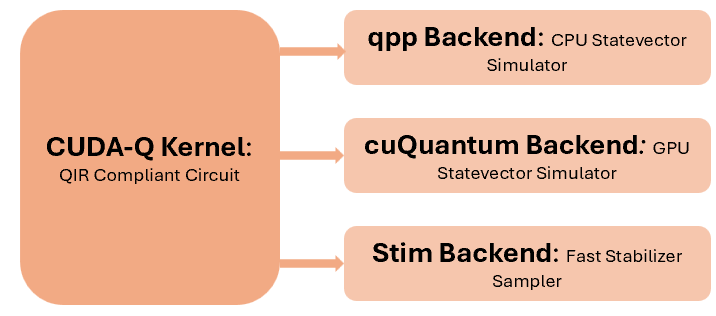}
        \caption{Backend execution options.}
        \label{fig:backend_execution}
    \end{subfigure}

   \caption{\textbf{Architecture of the Proposed Transpilation Framework.} (a) User OpenQASM~3.0 programs are parsed by the pyqasm frontend, transpiled into QIR-compliant CUDA-Q kernels, and executed by CUDA-Q. (b) QASM is lowered from AST to MLIR (classical-control + Quake dialects), then translated to LLVM IR/QIR and emitted as executable kernels. (c) CUDA-Q executes on qpp (CPU state-vector), cuQuantum (GPU state-vector), and Stim (stabilizer sampling) backends.}
    \label{fig:central_architecture}
    \vspace{-1em} 
    \label{fig:central_figure}
\end{figure*}

\section{Introduction}
The rapid evolution of Noisy Intermediate-Scale Quantum (NISQ) \cite{Preskill2018quantumcomputingin} devices has necessitated a parallel rapid advancement in quantum software infrastructure. As experimentalists have made an effort to approach the realization of fault-tolerant protocols, the software stack has fragmented into a diverse ecosystem of high-level Domain-Specific Languages (DSL), such as IBM's Qiskit, Google's Cirq, Microsoft's Azure Quantum, and Xanadu's Pennylane. While these frameworks facilitate circuit synthesis, the quantum software industry has increasingly coalesced around OpenQASM 3.0 \cite{cross22} as the standard intermediate representation (IR) for describing quantum programs. Unlike its predecessor, OpenQASM 2.0, OpenQASM 3.0 extends the specification to include pulse-level control, timing, and crucially the classical control flow (i.e., loops and conditionals) required for dynamic circuits and quantum error mitigation.

Despite this standardization, a significant gap remains in the tooling available to execute OpenQASM 3.0 programs efficiently. At the time of its release, OpenQASM 3.0 suffered from a lack of compiler infrastructure, often functioning merely as a parser rather than as a fully executable pipeline \cite{arulandu24}. This limitation is particularly acute when attempting to bridge standard quantum assembly with high-performance simulation backends. For instance, NVIDIA’s CUDA-Q \cite{Kim23} has emerged as a premier platform for GPU-accelerated quantum simulation, leveraging the Quake dialect of multi-level IR (MLIR) to achieve orders-of-magnitude speedups in state vector and tensor network simulations. However, the documentation for bridging standard OpenQASM 3.0 directly to CUDA-Q kernels remains sparse, creating a bottleneck for researchers attempting to leverage GPU acceleration for dynamic circuit validation.

In this work, we present a modular transpilation framework (Figure~\ref{fig:central_figure}) designed to bridge the gap between the expressive power of OpenQASM 3.0 and the execution performance of modern quantum backends. Our contribution is threefold:
\begin{enumerate}
    \item \textbf{Semantic Translation to High-Performance Kernels}: We introduce a Python-based transpiler that parses OpenQASM 3.0 source code—including complex gate modifiers, classical feedback, and control flow—and generates optimized both C++ and Python-based CUDA-Q kernels. This allows researchers to write standard assembly code and automatically gain access to NVIDIA’s GPU-accelerated simulation backend.
    \item \textbf{Interoperability with Qiskit}: We demonstrate the versatility of our framework by supporting transpilation to Qiskit circuit objects. This ensures that the same source code can be validated against the IBM mature software stack or executed on IBM hardware, promoting cross-platform verification.
    \item \textbf{Validation of Dynamic Circuits}: Addressing previous limitations in static circuit analysis, we provide a rigorous testing suite for dynamic quantum circuits. We validate classical feedforward and parameter passing mechanisms (essential for Variational Quantum Eigensolvers and error correction) by benchmarking generated kernels against theoretical expectations for conditional reset and teleportation protocols.
\end{enumerate}
We evaluate the performance of our transpiler by measuring compilation throughput and execution fidelity across a randomized set of Clifford circuits and standard algorithms. Our results indicate that this framework not only democratizes access to high-performance simulation but also serves as a critical tool for verifying the future work of quantum computing: dynamic, feedback-driven algorithm execution.

We begin our study by delineating the required background material.

\section{Background}
Before proceeding into the methodological implementations of our transpilation work, this section establishes the foundataional background required to understand the significance of our transpiler, specifically focusing on the shift from static to dynamic quantum execution and the necessity of efficient parameter handling for hybrid quantum algorithms.

\subsection{OpenQASM 3.0: From Description to Assembly}
Open Quantum Assembly Language (OpenQASM) \cite{cross17} has served as the de facto standard for describing quantum circuits since its inception. While OpenQASM 2.0 was designed primarily for representing straight-line quantum operations and measurements, it lacked the expressive power required for modern fault-tolerant protocols and hybrid algorithms. OpenQASM 3.0 represents a paradigm shift, extending the specification to include gate modifiers, timing operations, and classical control flow such as loops and conditionals.

Despite these advancements, the ecosystem initially suffered from a lack of robust tooling, offering little beyond a basic parser at the time of release \cite{arulandu24}. Although frameworks like Qiskit have begun to integrate OpenQASM 3.0 support, there remains a critical need for standalone transpilers capable of interpreting these high-level features—specifically dynamic flow control—and mapping them to performant backend IRs.

\subsection{The Transpilation Pipeline}
Quantum transpilation is the process of rewriting a quantum program into an equivalent topological implementation compatible with a specific target backend or simulator. This often involves gate decomposition (breaking down complex unitaries into a native gateset) and optimization.

As the quantum landscape fragments into various hardware topologies and assembly languages, the community has pushed toward unification via the Quantum Intermediate Representation (QIR) \cite{Geller22}, which is built on LLVM \cite{LLVM04}. A robust transpiler must bridge the gap between human-readable assembly (OpenQASM) and machine-optimizable IR (QIR/MLIR), enabling researchers to write agnostic code that executes efficiently across different platforms.

\subsection{CUDA-Q and High-Performance Simulation}
To mitigate the exponential cost of classical quantum simulation, NVIDIA introduced the cuQuantum SDK \cite{bayraktar23} and the CUDA-Q programming model. CUDA-Q represents quantum programs in MLIR, using the Quake dialect for quantum kernels. These kernels are compiled through LLVM and emitted as QIR, enabling efficient parallel execution on GPUs.

Bridging OpenQASM 3.0 to CUDA-Q is non-trivial because OpenQASM exposes explicit classical types and dynamic control flow, while Quake/C++ rely on a statically typed compilation pipeline. Our work addresses this translation gap by mapping OpenQASM 3.0 constructs—including classical logic—into executable CUDA-Q kernels.

\begin{figure}[t]
    \centering
    \includegraphics[width=0.85\linewidth]{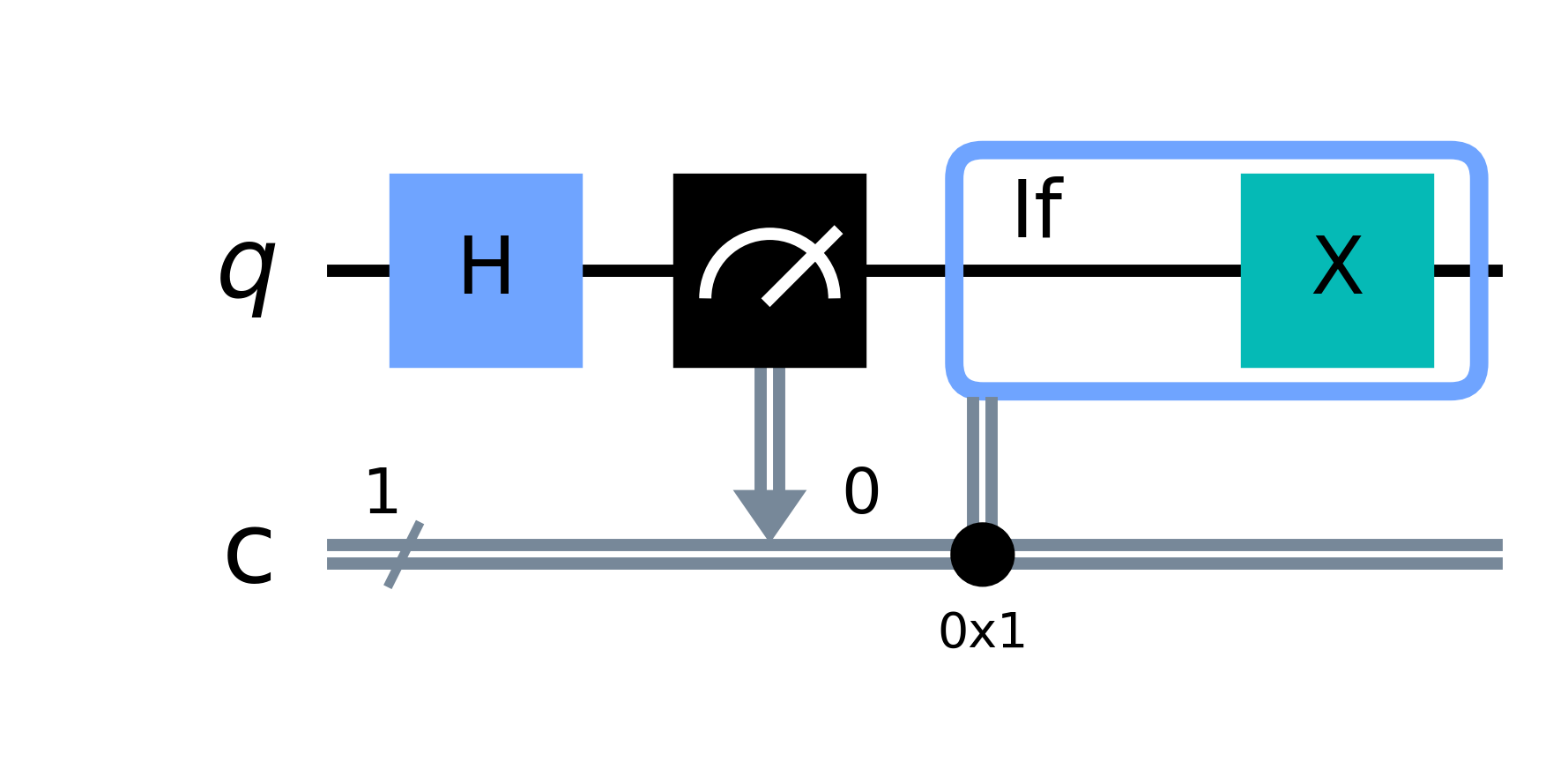}
    \caption{\textbf{Schematic of the Conditional Reset Protocol.}
    A qubit is prepared in $|+\rangle$ using a Hadamard gate, measured mid-circuit, and classically controlled: if the outcome is `1', a Pauli-X gate resets the state to $|0\rangle$.}
    \label{fig:dynamic_circuit}
\end{figure}

\subsubsection{Dynamic Circuit Testing}
OpenQASM 3.0 supports dynamic circuits, where gate sequences depend on mid-circuit measurement outcomes. This feature is fundamental to Quantum Error Correction (QEC) and other adaptive protocols. Figure~\ref{fig:dynamic_circuit} shows a conditional reset: measurement collapses the state, and real-time classical feedforward applies an $X$ gate only when needed, implementing a non-unitary correction primitive.

Compared to static unitary circuits, dynamic circuits can reduce quantum depth by shifting work to classical control, enabling capabilities such as long-range entanglement and teleportation-based connectivity \cite{B_umer_2024}. However, branching behavior makes verification harder and is poorly supported by many high-performance simulators. Our transpiler compiles OpenQASM 3.0 dynamic logic directly into CUDA-Q kernels, enabling scalable testing of adaptive workloads on CUDA-Q backends.

\subsection{Parameter Passing and VQE}
Variational Quantum Algorithms (VQAs) like VQE repeatedly execute a fixed ansatz $U(\vec{\theta})$ while updating parameters $\vec{\theta}$ \cite{Peruzzo_2014}. Recompiling the circuit at every iteration is prohibitively expensive; practical execution requires compiling once and supplying parameters at runtime.

OpenQASM 3.0 enables this with \texttt{input} variables, which are resolved during execution rather than compilation. A key transpiler benchmark is therefore correct and efficient parameter passing: mapping OpenQASM \texttt{input} variables to backend kernel arguments (e.g., CUDA-Q kernel parameters). This compile-once, run-many capability is essential for scalable VQE and other QML/chemistry workflows.

\section{Methodology}
Our transpilation framework is architected as a multi-stage compiler pipeline that transforms OpenQASM 3.0 source strings into executable CUDA-Q kernels. The pipeline consists of three distinct phases: syntactic parsing, semantic translation via Abstract Syntax Tree (AST) traversal, and kernel code generation.

\subsection{Parsing and Semantic Analysis}
We utilize the pyqasm library \cite{Gupta_PyQASM_Python_package_2025} to ingest OpenQASM 3.0 source code and generate a strongly-typed Abstract Syntax Tree (AST). Unlike simple regex-based parsers, pyqasm provides a rigorous representation of the language grammar, allowing us to handle complex nested structures.

Upon parsing, the system performs an initial pass to resolve symbol tables, tracking the scope and type of declared variables. This is critical for distinguishing between:

\begin{enumerate}
    \item \textbf{Compile-time constants}: Values known immediately (e.g., $\frac{pi}{2}$).
    \item \textbf{Runtime parameters}: Variables declared with \texttt{input} (e.g., angles for VQE).
    \item \textbf{Classical registers}: Variables storing measurement results (e.g., \texttt{bit c}).
\end{enumerate}

\subsection{Kernel Generation via AST Visitor}
The core translation logic employs the visitor design pattern. A specialized CUDAQVisitor class traverses the OpenQASM AST depth-first. For every node encountered (e.g., QuantumGate, Measure, IfStatement), the visitor emits the corresponding CUDA-Q Python API calls to construct the kernel.

Standard unitary operations are mapped directly to their CUDA-Q equivalents (e.g., \texttt{h q[0]}  $\rightarrow$ \texttt{kernel.h(q[0])}). However, OpenQASM 3.0 gate modifiers—specifically \texttt{ctrl} (controlled) and inv (adjoint)—require special handling.
When the visitor encounters a modified gate, it does not emit the instruction directly. Instead, it encapsulates the target operation into a temporary sub-kernel or lambda function and utilizes CUDA-Q’s functional API (e.g., \texttt{cudaq.control(sub\_kernel, ...)} or \texttt{cudaq.adjoint(...)}) to apply the modifier dynamically.

\subsection{Transpiling Dynamic Control Flow}
A significant contribution of this work is the support for dynamic circuits, which requires mapping OpenQASM's explicit control flow to CUDA-Q's kernel-based logic.

When the visitor encounters an \texttt{if} statement node, the following translation occurs:

\begin{enumerate}
    \item \textbf{Condition Extraction}: The visitor identifies the classical bit or register serving as the predicate (e.g., \texttt{if (c == 1)}).
    \item \textbf{Body Encapsulation}: The quantum operations inside the \texttt{if} block are captured into a Python callable (a closure) rather than being executed immediately.
    \item \textbf{Conditional Emission}: The transpiler emits a \texttt{kernel.c\_if(measurement\_result, callable}) instruction. This defers execution to the hardware or simulator's control engine, enabling true classical feedforward rather than pre-calculated branching.
\end{enumerate}

\subsection{Handling Runtime Parameters (VQE Support)}
To support variational algorithms efficiently, we implemented a mechanism to map OpenQASM \texttt{input} declarations to CUDA-Q kernel arguments. When the parser detects an \texttt{InputDeclaration} (e.g., \texttt{input float[2] theta;}), the transpiler modifies the signature of the generated kernel to accept arguments of the corresponding type (List[float]). References to these variables within gate operations (e.g., \texttt{rx(theta[0]) q[0]}) are resolved not as float literals, but as symbolic references to the kernel arguments. This allows the resulting CUDA-Q kernel to be compiled once and executed repeatedly with different numerical inputs, avoiding the Just-In-Time (JIT) recompilation overhead that plagues naive transpilation approaches.

\subsection{Validation Framework}
To verify the correctness of dynamic features, we moved beyond the static statevector comparison methods used in previous works. We developed a Dynamic Validation Suite consisting of two primary test vectors:
\begin{itemize}
    \item \textbf{Deterministic Control Flow}: Circuits such as "Conditional Reset," where non-deterministic measurement outcomes are corrected by feedforward logic to produce a deterministic final state (e.g., always $|0\rangle$).
    \item \textbf{Algorithmic Parity}: Verification of parameterized circuits (VQE ansatzes) by sweeping input parameters and asserting that the energy expectation values match theoretical predictions within a defined tolerance.
\end{itemize}

\section{Experiments}

\subsection{Implementation}
\subsubsection{Implementation and Experimental Setup}
The proposed transpilation framework was implemented in Python 3.10, utilizing the pyqasm library  for OpenQASM 3.0 parsing and semantic analysis. The core translation logic was built using the Visitor pattern to map the OpenQASM Abstract Syntax Tree (AST) to the NVIDIA CUDA-Q Python API (Version 0.7.1). To enable cross-platform validation, we also integrated Qiskit 1.0  as a secondary transpilation target.

All experiments were conducted in a high-performance simulation environment.  The CUDA-Q backend was configured to use the qpp-cpu target for the CPU-based state-vector simulation and the nvidia backend with fp32 precision for GPU-accelerated runs. These correspond to the Q++ CPU simulator and the cuQuantum single-GPU simulator, respectively.

\subsubsection{Benchmark Suites}
To rigorously evaluate the transpiler’s capabilities, we designed three distinct benchmark suites covering static, dynamic, and hybrid quantum execution.

\begin{enumerate}
    \item \textbf{Static Circuit Verification (Clifford and Algorithms)}: Following standard validation protocols, we utilized a randomized testing suite generating Clifford circuits of varying depth (range: 10–100 gates) and qubit count (range: 2–20). Additionally, we included standard algorithms such as the Quantum Fourier Transform (QFT)  and Bernstein-Vazirani. Correctness was verified by comparing the final state vector of the CUDA-Q kernel against the output of Qiskit’s \texttt{StatevectorSimulator}, ensuring fidelity up to global phase.
    \item \textbf{Dynamic Circuit Validation (Control Flow)}: To address the limitations of static analysis, we implemented a deterministic functional test suite for dynamic features. This included:
    \begin{itemize}
        \item \textbf{Conditional Reset}: A single-qubit test preparing $|-\rangle$, measuring, and conditionally applying $X$ to force a $|0\rangle$ state. Success was defined as measuring $|0\rangle$ with $>99.9\%$ probability across 1000 shots.
        \item \textbf{Quantum Teleportation}: A three-qubit protocol utilizing mid-circuit measurement and classical feedforward to teleport a state. Correctness was validated by uncomputing the teleported state and asserting a deterministic return to $|0\rangle$.
    \end{itemize}
    \item \textbf{Hybrid VQE Parameterization:} To validate the efficiency of the \texttt{input} variable mapping, we constructed a Hardware Efficient Ansatz (HEA) typical of Variational Quantum Eigensolver (VQE) tasks. The experiment simulated a VQE optimization loop by passing 50 distinct sets of parameters to a single compiled kernel. We measured both the execution time and the correctness of the energy expectation values $\langle H \rangle$ for a target Hamiltonian ($H = Z_0 Z_1$), verifying that the transpiler correctly mapped classical float arrays to kernel arguments without triggering recompilation.
\end{enumerate}

\subsubsection{Software Architecture}
The transpiler is architected as a modular Python package, leveraging the \texttt{pyqasm} library for robust OpenQASM 3.0 parsing. The core translation engine is implemented via the Visitor pattern in a python class called \texttt{OpenQASMToCUDA}. This class inherits from the standard AST visitor and overrides node-specific methods (e.g., \texttt{visit\_QuantumGate}, \texttt{visit\_IfStatement}) to emit the corresponding CUDA-Q Python API calls.

To handle dynamic control flow, the \texttt{visit\_IfStatement} method encapsulates the conditional body into a Python callable and invokes \texttt{cudaq.make\_kernel().c\_if()}. For variational algorithms, the \texttt{visit\_InputDeclaration} method maps OpenQASM \texttt{input} variables to typed CUDA-Q kernel arguments (e.g., \texttt{float}), enabling the compilation of parametric kernels that avoid Just-In-Time (JIT) recompilation overhead.

\subsubsection{Experimental Setup}
All experiments were conducted within a Docker container to demonstrate the use of the transpilation process for generalizability for users who lack CUDA-capable GPU hardware and can still generate and (CPU) simulate CUDA-Q programs inside a Docker container. We utilized the \texttt{qpp-cpu} backend for state vector simulations. The performance of the transpiler was evaluated against a baseline of standard Qiskit compilation and naive JIT recompilation strategies.

\subsection{Results}
We evaluated the performance and correctness of the transpiler across three distinct test suites: static circuit fidelity, dynamic control flow validation, and parameter passing tests for hybrid algorithms (i.e., VQE) efficiency. All tests were executed using the \texttt{nvidia-fp32} simulator backend.

We validated the transpiler against a suite of randomized Clifford circuits of varying depth ($d \in [10, 100]$) and width ($n \in [2, 20]$). We compared the output statevectors of the generated CUDA-Q kernels against Qiskit's \texttt{AerSimulator}. Our transpiler achieved 100\% semantic correctness across all static test cases, successfully mapping standard gates, custom gate definitions, and modifiers (e.g., \texttt{ctrl}, \texttt{inv}) to their Quake IR equivalents.

\subsection{Dynamic Control Flow Validation}
To assess the support for OpenQASM 3.0's dynamic features, we utilized a "Conditional Reset" protocol 
This test prepares a qubit in the $|-\rangle$ state, measures it, and conditionally applies an $X$ gate to force the state to $|0\rangle$. 

The transpiler correctly mapped the OpenQASM \texttt{IfStatement} AST nodes to CUDA-Q's \texttt{c\_if} construct. In stochastic simulations ($10^4$ shots) with a noisy simulator, the generated kernel produced the target state $|0\rangle$ with probability $P > 0.999$, confirming the correct execution of classical feedforward logic.

\subsection{Hybrid Compilation Efficiency (VQE)}
To demonstrate the advantage of parametric compilation for variational algorithms, we benchmarked the mapping of OpenQASM \texttt{input} variables. In standard transpilation workflows, updating a parameter $\theta$ requires string manipulation and re-parsing of the entire circuit (JIT recompilation).

Our transpiler identifies \texttt{input} declarations and transforms them into typed arguments in the CUDA-Q kernel signature. 
This allows the kernel to be compiled once and executed repeatedly with new numerical values. We observed a significant reduction in loop latency for a 50-iteration VQE simulation compared to the naive recompilation approach.

\section{Discussion}
The successful validation of conditional reset and teleportation benchmarks demonstrates that our transpiler correctly supports dynamic control flow, including mid-circuit–conditioned if statements that are required for Quantum Error Correction and are difficult to realize in many existing compilers. By mapping OpenQASM 3.0 \texttt{IfStatement} nodes directly to CUDA‑Q’s \texttt{c\_if} construct, the framework enables practical testing of fault-tolerant primitives such as repeat‑until‑success routines and magic state distillation on high-performance GPU simulators before deployment to hardware. At the same time, a Dockerized simulation environment broadens accessibility by allowing users to verify OpenQASM 3.0 semantics on standard CPU-only systems, supporting the development of dynamic algorithms in resource-constrained settings while preserving a clear migration path to multi-GPU configurations when such resources become available.

\section{Conclusion and Future Directions}
While our current implementation targets the abstract gate model, future work will extend the transpiler to interpret OpenQASM 3.0 pulse-level instructions and map them to CUDA‑Q’s analog control interfaces, enabling advanced error-mitigation protocols that require precise timing and pulse shaping. We also plan to adopt the Quantum Intermediate Representation (QIR) as a primary emission target to decouple language analysis from backend execution, thereby achieving interoperability with a broader ecosystem of hardware platforms beyond NVIDIA GPUs. Finally, by leveraging cuQuantum’s density-matrix and tensor-network capabilities to model depolarization and readout errors during dynamic control, we aim to move beyond ideal state-vector simulations and rigorously assess quantum error-correcting codes under realistic decoherence constraints.

\bibliographystyle{IEEEtran}
\bibliography{ref1}

\end{document}